\documentclass[runningheads]{llncs}
\usepackage{amsmath,amssymb,xcolor}
\usepackage[T1]{fontenc}
\usepackage{graphicx,booktabs,url}
\usepackage{subcaption}
\usepackage{multirow}

\begin{document}
\title{EduExplain: A Benchmark for Explanation-Set Retrieval and Storyboard Construction in Educational Explanation}

\author{Aryan Ahmed \and Riti Verma \and Sanskar Singh \and Maya Ramanath}
\institute{IIT Delhi, New Delhi, India\\
\email{ahmedaryan001@gmail.com}\\
\email{riti.iitd.cse@gmail.com}\\
\email{ssanskar2705@gmail.com}\\
\email{ramanath@cse.iitd.ac.in}}

\maketitle

\begin{abstract}
Generating an educational explanation for a learner's query is not a single retrieval step but a pipeline: a system must first \emph{retrieve} a compact set of complementary resources---passages, figures, examples, activities---that jointly support an explanation, and then \emph{structure} that set into an ordered, timed, role-assigned plan before it can be rendered as a lesson or video. Existing retrieval benchmarks evaluate neither stage at the level these applications require: they score individual documents rather than explanation sets, and they provide no ground truth for how a set should be organised into an explanation. We present EduExplain, a benchmark that provides human-validated ground truth for both stages. Built from secondary-level science textbooks, it comprises 6{,}921 typed, multimodal resources and 2{,}123 concepts, and pairs each of 321 learner-oriented queries with one or more human-validated explanation sets (retrieval ground truth) and 741 reference storyboards (structuring ground truth). The benchmark exposes characteristics absent from prior datasets: heterogeneous multimodal retrieval units, multiple valid explanation sets per query, and plans whose quality depends on pedagogical role composition, grounding, and timing rather than relevance alone. We describe an LLM-assisted, human-validated construction pipeline, report inter-annotator agreement for both layers, and establish reference baselines for each stage---lexical, dense, multimodal, and cross-encoder retrieval paired with four selection strategies, and free-form and fixed-template storyboard construction. Our results show that first-stage retrieval quality dominates retrieval-stage performance, that off-the-shelf diversity-based selection underperforms simple top-ranked selection, and that unconstrained and rigidly templated structuring both fall short of the reference plans---establishing EduExplain as a testbed for set-level, pedagogically-aware educational explanation. The benchmark is available at \url{https://github.com/data-iitd/eduexplain}.

\keywords{Explanation-set retrieval \and Educational benchmarks \and Multimodal retrieval \and Storyboard construction.}
\end{abstract}

\section{Introduction}

Recent advances in large language models, multimodal foundation models, and generative AI have enabled a new class of educational applications, including conversational tutoring systems, automatic lesson generation, curriculum-aware question answering, and educational video generation. These systems increasingly rely on retrieval to ground generated content in authoritative educational resources. In practice, however, educational content is rarely organized as a collection of standalone documents. Instead, textbooks contain a rich mixture of explanatory passages, worked examples, diagrams, and activities that collectively contribute to learning a concept.

Producing an explanation from such material is a multi-stage process. Consider a learner asking ``Why do plants need sunlight?'' A conventional retrieval system may return an entire chapter on photosynthesis or several highly relevant passages. While useful for information access, this is not yet an explanation: a tutoring response or short instructional video needs only a small subset of the material---say, a definition, a process diagram, and an illustrative example---and it must decide how to present that subset: what to show first, which resource serves which pedagogical function, and how long to spend on each. We therefore view explanation generation as a pipeline with three stages: \emph{retrieve} a compact set of complementary resources, \emph{structure} that set into an ordered, timed, role-assigned plan, and \emph{render} the plan as narrated output. The first two stages determine whether the explanation is complete and well-organised; the third realises it. EduExplain provides evaluable ground truth for the first two stages and treats rendering as downstream and out of scope.

The first stage, which we call \emph{explanation-set retrieval}, departs from conventional retrieval in that the target is not a single ranked item but a coherent \emph{set}. Multiple valid explanation sets may exist for the same query: photosynthesis may be explained using a definition and a diagram, or through an activity and a worked example, and both may be pedagogically sound despite drawing on different resources. The retrieval target is therefore a set of resources that together suffice for an explanation, and several such sets may be acceptable.

The second stage, which we call \emph{grounded storyboard construction}, takes a retrieved set and organises it. A \emph{storyboard} is an ordered sequence of segments, each assigning a pedagogical role (define, explain a mechanism, give an example, check understanding), grounding that role in specific resources from the set, and allocating a share of a time budget. A plan that presents an example before the definition it depends on, or that exceeds its budget, is a structuring failure that set-level retrieval metrics cannot detect. This stage has its own ground truth and its own notion of quality---role composition, ordering, grounding, and timing---distinct from retrieval relevance.

Existing retrieval benchmarks are designed for neither stage. Document and passage benchmarks assume the retrieval unit is a whole document and that relevance can be assessed independently per item, so they neither evaluate explanation sets nor provide any structuring ground truth. Evidence-retrieval and multi-fact benchmarks evaluate text-only fact sets for verification, not multimodal resource sets organised for pedagogy. EduExplain fills this gap by evaluating both stages directly: it measures whether a system retrieves a compact multimodal set sufficient for an explanation, and whether it can organise such a set into a plan that matches human-authored references.

EduExplain is built from secondary-level science textbooks. It comprises 6{,}921 fine-grained, typed resources---4{,}594 explanatory passages, 1{,}717 figures, 495 worked examples, and 115 activities---annotated with 2{,}123 concepts derived from chapter and section topics. For each of 321 learner-oriented queries, spanning seven educational intents and three surface forms, the benchmark provides one or more human-validated explanation sets and, as a companion layer, 741 human-reviewed reference storyboards. Both layers are produced by an LLM-assisted pipeline and validated by human annotators, and we report inter-annotator agreement for each.

Our contributions are: (i) we frame educational explanation as a staged pipeline and define its two evaluable stages---explanation-set retrieval and grounded storyboard construction---as the object of a benchmark; (ii) we release EduExplain, which provides human-validated ground truth for both stages: 321 queries with explanation sets over 6{,}921 typed multimodal resources, and 741 reference storyboards; (iii) an LLM-assisted, human-validated construction pipeline with reported inter-annotator agreement for both layers; (iv) stage-appropriate evaluation metrics---exact explanation-set recovery and conceptual coverage for retrieval, and role composition, grounding, coverage, and timing for structuring; and (v) reference baselines for each stage, showing that first-stage retrieval quality dominates retrieval performance, that diversity-based selection underperforms top-ranked selection, and that free-form and fixed-template structuring both fall short of the reference plans. We hope EduExplain provides a foundation for future work on retrieval, structuring, and generation of educational explanations.

\section{Problem Formulation}

We formalize the two stages EduExplain evaluates. Both operate over a shared corpus of educational resources and are motivated by applications that answer a learner query by constructing a concise explanation from textbook-derived material: \emph{explanation-set retrieval} selects a compact set of resources sufficient for an explanation, and \emph{grounded storyboard construction} organises such a set into an ordered, timed, role-assigned plan. Unlike standard document or passage retrieval, neither stage has a single ranked item as its target.

\subsection{Educational Resources}

Let $C$ denote a corpus of educational resources extracted from textbooks. A resource is an atomic instructional unit that can be independently retrieved and used during explanation construction. Each resource $r \in C$ has a pedagogical type $\mathrm{type}(r) \in \mathcal{T} = \{\text{passage}, \text{figure}, \text{example}, \text{activity}\}$ and a modality $\mathrm{mod}(r) \in \{\text{text}, \text{image}\}$, where figures are the image modality and all other types are text. Every resource carries a unique identifier, source textbook, chapter and section metadata, a textual representation (its own text, or a generated description for figures), a concise summary, and concept annotations $\mathrm{Con}(r) \subseteq \mathcal{V}$ drawn from a controlled concept vocabulary $\mathcal{V}$. Associating a textual representation with every resource, figures included, preserves the multimodal origin of each resource while providing the uniform textual form needed for retrieval and annotation.

\subsection{Queries}

A query $q$ represents a learner information need. Queries may ask for a basic explanation, a causal or procedural explanation, a misconception clarification, an example, a visual explanation, a revision summary, or practice material. A query such as ``Why do plants need sunlight?'' is not merely a keyword lookup for the photosynthesis chapter: it expresses a focused explanatory need that can be satisfied by a small subset of the chapter's resources rather than the chapter as a whole.

\subsection{Explanation Sets}

Given a learner query $q$, the first-stage goal is to retrieve a set of resources $R_q = \{r_1, \ldots, r_k\}$, $r_i \in C$, that collectively supports a concise explanation of $q$. We call such a set an \emph{explanation set}. An explanation set is not required to contain a fixed combination of resource types: a short textual explanation suffices for some queries, while others benefit from a figure, an example, or an activity. The task therefore does not reduce to independently retrieving the top passage, the top figure, and the top example; the system must identify a compact set of resources sufficient for the specific query.

For each benchmark query, EduExplain provides one or more human-validated gold explanation sets $G_q = \{G_q^{(1)}, \ldots, G_q^{(m)}\}$, where each $G_q^{(i)}$ is a set of resource identifiers judged sufficient to support a concise explanation of $q$. Multiple gold sets are included because a concept may be explained in several valid ways---for example, photosynthesis may be explained using a passage and a process figure, or through an activity and a worked example.

\subsection{Grounded Storyboard Construction}
\label{sec:gsc}

The second stage takes a query $q$ and an explanation set $R_q$ as fixed input and organises the set into a presentation plan. Each segment carries a \emph{pedagogical role} describing its explanatory function rather than its resource type, drawn from
\[
\begin{aligned}
\mathcal{R} = \{\, & \textsc{hook}, \textsc{context}, \textsc{definition}, \textsc{mechanism}, \\
                   & \textsc{comparison}, \textsc{example}, \textsc{visual}, \textsc{activity}, \\
                   & \textsc{check}, \textsc{misconception}, \textsc{summary} \,\}.
\end{aligned}
\]
A \emph{storyboard} for $q$ over $R_q$ is an ordered sequence of segments $B_q = (\sigma_1, \ldots, \sigma_L)$, where each segment $\sigma_\ell = \langle \rho_\ell, S_\ell, \tau_\ell \rangle$ carries a role $\rho_\ell \in \mathcal{R}$, a non-empty grounding subset $S_\ell \subseteq R_q$ of resources, and a planned duration $\tau_\ell$, with total duration $T(B_q) = \sum_\ell \tau_\ell$. Grounded storyboard construction is the task of producing, from $q$ and $R_q$ under a target duration $T^{\mathrm{tgt}}$, a storyboard that maximises explanatory quality subject to resource grounding ($\emptyset \neq S_\ell \subseteq R_q$ for every $\ell$) and the time budget ($|T(B_q) - T^{\mathrm{tgt}}| \le \delta$). An example presented before the definition it depends on, or a plan that exceeds its budget, is a structuring failure that set-level retrieval metrics cannot detect. Because a query admits several valid storyboards, EduExplain provides one or more human-validated reference storyboards $B_q^\star = \{B_q^{\star(1)}, \ldots\}$ per query, and quality is credited against the best-matching reference.

\section{Related Work}

EduExplain lies at the intersection of multimodal question answering, evidence retrieval, multimodal document retrieval, and educational content generation. Each addresses a related problem, but none evaluates both the retrieval of a compact multimodal explanation set and its organisation into a pedagogical plan.

\subsection{Multimodal Question Answering}

Benchmarks such as VisDoM~\cite{visdom} and MMLongBench~\cite{mmlongbench} evaluate question answering over visually rich documents where the relevant information is distributed across text, figures, tables, and charts. Closest to our domain, TQA~\cite{kembhavi} targets science education, pairing multimodal middle-school lessons with multiple-choice and true/false questions that require joint reasoning over text and diagrams. These benchmarks evaluate the correctness of the final answer, with retrieval serving only as an intermediate component of a QA pipeline. EduExplain instead evaluates the retrieved set itself: rather than asking whether a model can answer once given sufficient context, it asks whether a system can retrieve a compact collection of complementary resources that together support an explanation. A resource set useful for explanation is not the same as an evidence set sufficient for answering a factual question, and TQA in particular neither decomposes lessons into typed pedagogical units nor admits alternative valid explanatory compositions.

\subsection{Evidence Retrieval and Multi-Fact Explanation Retrieval}

Claim-verification benchmarks such as FEVER~\cite{fever} and SciFact~\cite{scifact} were among the first to evaluate retrieval at the level of evidence sets rather than individual documents, but remain text-only and focused on evidential sufficiency for verification rather than explanatory sufficiency for education. The closest prior work is the WorldTree corpus~\cite{worldtree,worldtreev2} and the associated TextGraphs explanation-regeneration tasks~\cite{textgraphs}, which require retrieving the set of facts that jointly explain a science exam answer and have framed multi-fact explanation retrieval as an IR objective~\cite{valentino}. EduExplain differs in three respects. First, WorldTree's units are curated, text-only facts from a semi-structured knowledge base, whereas EduExplain's are heterogeneous multimodal resources---passages, figures, examples, and activities---extracted from unstructured textbook chapters. Second, WorldTree explanations are assembled by lexical and semantic overlap along a single inference chain toward a fixed answer, whereas EduExplain requires composing resources that play complementary pedagogical roles (a definition, an illustration, a worked example) for an open-ended query, where textual overlap is not the organizing principle. Third, WorldTree gives each question a single gold explanation, while EduExplain encodes multiple independently valid explanation sets per query.

\subsection{Multimodal and Diversified Retrieval}

MMDocIR~\cite{mmdocir} makes retrieval a first-class evaluation target over multimodal documents, but its target is an individually relevant page or layout element rather than a coherent set: a single relevant figure may be insufficient when the query also needs prerequisite information, a worked example, or an assessment item. EduExplain is also related to diversified retrieval---MMR~\cite{mmr}, Determinantal Point Processes~\cite{dpp}, and graph-based selection---which reduce redundancy within a ranked list. We include such methods among our reference selection baselines; because the resources of a good explanation are often semantically close, pairwise-diversity objectives are a natural but imperfect fit for explanation-set construction.

\subsection{Educational Content Generation}

A recent line of work generates educational artifacts end to end. TheoremExplainAgent~\cite{theorem} produces theorem-explanation videos; EduStory~\cite{edustory} introduces pedagogical state modelling and structured multi-shot control with storyboard annotations; CourseBlueprint~\cite{courseblueprint} uses validated intermediate representations, prerequisite-aware scaffolding, and a fixed narration contract over retrieved slides; and EduIllustrate~\cite{eduillustrate} evaluates interleaved text--diagram explanations for K--12 problems. These systems make instructional planning explicit but couple retrieval, structuring, and rendering within a single model, so the quality of the retrieved set and of its organisation cannot be isolated. EduExplain instead provides stage-wise ground truth---human-validated explanation sets and reference storyboards---so that retrieval and structuring can be evaluated independently of each other and of rendering.

\section{Benchmark Construction}

\begin{figure*}[t]
\centering
\includegraphics[width=\textwidth]{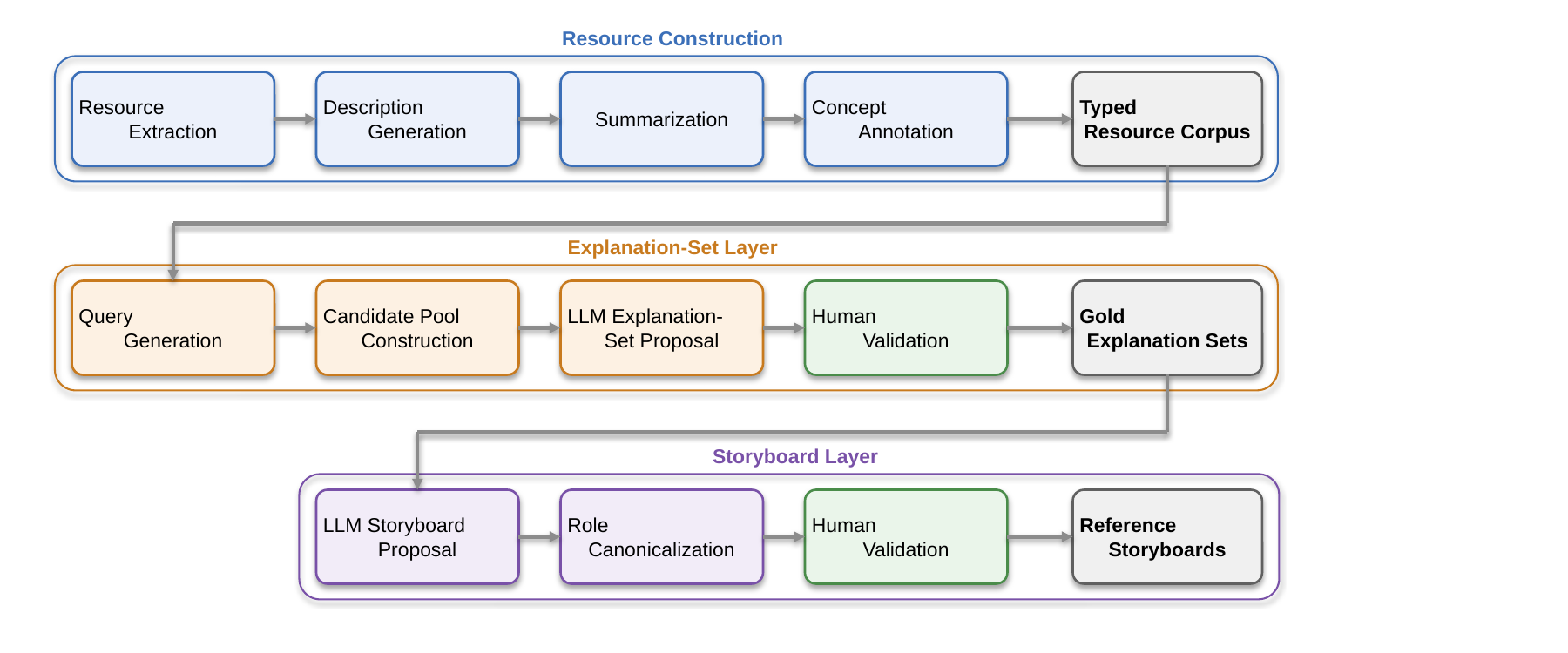}
\caption{EduExplain construction pipeline. Resource construction yields a typed, multimodal resource corpus. The explanation-set layer generates and human-validates gold explanation sets over that corpus; the storyboard layer then organises each validated explanation set into reference storyboards, with its own human-validation stage. The benchmark releases both validated layers.}
\label{fig:pipeline}
\end{figure*}

Figure~\ref{fig:pipeline} summarizes the two-layer construction pipeline. Starting from secondary-level science textbooks from different Indian school boards, we extract fine-grained educational resources and generate textual representations and concept annotations, producing a typed multimodal corpus. From this corpus we construct learner-oriented queries, propose candidate explanation sets with a large language model, and human-validate them to obtain the gold explanation sets. Each validated explanation set is then organised into a reference storyboard---an ordered, timed, role-assigned plan---which undergoes a second human-validation stage. Both layers are released.

\subsection{Resource Extraction and Annotation}

EduExplain is constructed from secondary-level science textbooks. Each chapter is decomposed into fine-grained, typed educational resources that serve as the retrieval units: explanatory passages, figures, worked examples, and activities. Unlike conventional retrieval benchmarks that index whole documents or passages, EduExplain represents the heterogeneous instructional material found in textbooks. Each resource is assigned a unique identifier together with its textbook, chapter, section, and subsection metadata.

To provide a common representation across modalities, every resource is associated with a textual description. For textual resources, the original text is retained together with a concise LLM-generated summary. For figures, a vision-language model generates an educational description that captures the concepts illustrated rather than merely the visual appearance. These descriptions provide a unified textual representation used throughout benchmark construction. Throughout construction we use Qwen2.5-7B-Instruct as the language model (for summarization, concept annotation, query generation, and gold-set proposal) and Qwen2.5-VL-7B-Instruct as the vision-language model for figure descriptions.

Each resource is then annotated with one or more educational concepts using a large language model, and the annotations are validated by human annotators. Concept labels are drawn from a controlled vocabulary derived from textbook chapter and section topics, so that a resource may contribute to multiple concepts while a concept may be represented by multiple heterogeneous resources. The final benchmark stores, for every resource, its identifier, resource type, source location, original content (or image reference), textual summary, and concept annotations.

\subsection{Query Generation}

\begin{table*}[t]
\centering
\small
\begin{tabular}{p{2.5cm} p{5.5cm} p{5cm}}
\hline
\textbf{Query Type} & \textbf{Purpose} & \textbf{Example Query} \\
\hline

Basic Explanation &
Requests a direct explanation of the concept. &
Explain photosynthesis. \\

Why/How &
Seeks causal or procedural understanding. &
Why do plants need sunlight? \\

Misconception &
Reflects a common misunderstanding or incorrect belief. &
Do plants get food from the soil? \\

Example-Based &
Requests an illustrative example to aid understanding. &
Explain photosynthesis using a daily-life example. \\

Visual &
Requests a diagrammatic or visual explanation. &
Show me how photosynthesis happens in a leaf. \\

Revision &
Requests a concise summary or recap of the concept. &
Give me a quick recap of photosynthesis. \\

Practice &
Requests exercises, assessment questions, or self-testing material. &
Give me a question to test my understanding of photosynthesis. \\

\hline
\end{tabular}
\caption{Query categories used during benchmark construction. For each concept, the language model is instructed to generate multiple learner-oriented queries corresponding to each category. The example queries in this table are for the concept "Photosynthesis".}
\label{tab:querytypes}
\end{table*}

\begin{table}[t]
\small
\centering
\begin{tabular}{@{}lrrrrrrr@{}}
\toprule
& & & \multicolumn{4}{c}{Gold sets per query ($m$)} & \\
\cmidrule(lr){4-7}
Query type & Queries & \% & Mean & $m{=}1$ & $m{=}2$ & $m{=}3$ & Mean $|G|$ \\
\midrule
Example       &  66 & 20.6 & 3.00 &  0 & 0 &  66 & 2.54 \\
Revision      &  61 & 19.0 & 3.00 &  0 & 0 &  61 & 2.72 \\
Practice      &  57 & 17.8 & 1.18 &  52 & 0 &  5 & 2.67 \\
Why/How       &  54 & 16.8 & 2.80 &  2 & 7 &  45 & 2.54 \\
Explanation   &  32 & 10.0 & 2.59 &  2 & 9 &  21 & 2.30 \\
Visual        &  27 &  8.4 & 1.07 & 25 & 2 &   0 & 1.59 \\
Misconception &  24 &  7.5 & 1.25 & 18 & 6 &   0 & 1.88 \\
\midrule
Total         & 321 & 100.0 & 2.31 & 99 & 24 & 198 & 2.44 \\
\bottomrule
\end{tabular}
\caption{Composition of EduExplain by query category. $m$ is the number of alternative gold explanation sets per query; $|G|$ is the number of resources in a single gold set, averaged per query then over the category. Totals: 321 queries, 741 gold sets; no gold set exceeds three resources.}
\label{tab:composition}
\end{table}
The benchmark evaluates retrieval systems that support educational explanation rather than conventional keyword search, so queries are generated to reflect realistic learner information needs. For every annotated concept, a large language model generates multiple learner-oriented queries spanning the seven educational intents in Table~\ref{tab:querytypes}: Explanation, Why/How, Misconception, Example-Based, Visual, Revision, and Practice. The model is prompted with the concept and its surrounding textbook context and instructed to produce natural-language questions appropriate for the target educational level. All generated queries are manually reviewed for grammatical correctness, educational relevance, and consistency with the intended category. Queries are generated at two levels of conceptual scope: from a single concept, and from groups of two or three concepts that co-occur within the same chapter, so that a query may require resources spanning several related concepts; we refer to this scope as the query's \emph{concept count}. Each resulting information need is additionally authored in three surface forms that share the same concept, intent, and gold explanation sets but differ in wording: a shorter search-like query, a conversational turn, and a fuller learning-oriented question, separating \emph{what} a learner asks from \emph{how} they phrase it.

\subsection{Candidate Resource Pool Construction}

Generating explanation sets directly from the entire resource collection is impractical, since a language model cannot process thousands of resources simultaneously. We therefore construct a candidate pool for each query from four complementary sources:
\begin{enumerate}
    \item resources annotated with the query concept,
    \item resources from the same textbook section or subsection,
    \item resources from the same chapter, and
    \item top-ranked results returned by standard retrieval methods (BM25 and dense retrieval with \texttt{bge-base-en-v1.5}).
\end{enumerate}
The first three sources exploit textbook structure and concept annotations; the fourth introduces relevant resources that may not have been explicitly linked to the concept. The union is deduplicated and ranked to produce a candidate pool of approximately 50--100 resources, which serves as the search space for explanation-set generation.

\subsection{Explanation Set Generation}

For each learner query, a large language model receives the candidate pool together with the textual summaries and metadata of its resources -- the full resource content is not used at this stage, so all modalities are presented through a unified textual representation. The model is instructed to identify the smallest collection of complementary resources sufficient to support a concise educational explanation while avoiding unnecessary redundancy. Because a concept may be explained in several valid ways, the model generates multiple alternative explanation sets per query, each consisting of a set of resource identifiers together with a short rationale describing the role of the selected resources. These LLM-generated sets are treated solely as annotation proposals rather than benchmark labels.

\subsection{Explanation-Set Validation}
\label{sec:validation}

\begin{table}[t]
\centering
\small
\begin{tabular}{lcc}
\toprule
Criterion & Krippendorff's $\alpha$ & Raw agreement \\
\midrule
Explanation completeness & 0.64 & 83\% \\
Resource relevance       & 0.44 & 84\% \\
Resource redundancy      & 0.28 & 75\% \\
\midrule
Query naturalness        & $\dagger$ & 96\% \\
Category correctness     & $\dagger$ & 94\% \\
Grade-level consistency  & $\dagger$ & 99\% \\
\bottomrule
\end{tabular}
\caption{Inter-annotator agreement on explanation-set validation, over a stratified sample, reported as Krippendorff's ordinal $\alpha$ and raw agreement. For the three strongly skewed criteria ($\dagger$) one category dominates, so $\alpha$ is deflated by the low-variance paradox and raw agreement is the faithful indicator; agreement is substantial on completeness and relevance, the properties the gold sets must satisfy.}
\label{tab:validation}
\end{table}

Validation proceeds in two stages. First, an automatic script removes generations that fail structural checks---malformed output that does not parse into the required format, or gold sets referencing hallucinated resource identifiers absent from the corpus. This deterministic filter requires no human judgment. Second, the surviving candidate explanation sets are reviewed on a stratified sample by a pool of four annotators, each set independently judged by two, with roughly one in five proposals rejected---so the human stage performs substantive filtering rather than rubber-stamping. Annotators score six criteria: \emph{completeness} (do the resources together fully answer the query), \emph{resource relevance} (does every resource genuinely help), \emph{non-redundancy} (does each resource earn its place), \emph{query naturalness} (would a real student ask this), \emph{category correctness} (does the query match its assigned intent), and \emph{grade-level consistency} (is the material appropriate to the target grade). We report inter-annotator agreement as both raw agreement and Krippendorff's ordinal $\alpha$ (Table~\ref{tab:validation}). Agreement is substantial on the criteria that matter most for a gold explanation set---completeness and relevance---establishing that the validated gold sets are reliable on the properties our metrics depend on.

\subsection{Storyboard Layer}
\label{sec:storyboard_layer}

The second layer extends each gold explanation set into a reference storyboard---an ordered, timed, role-assigned plan (Section~\ref{sec:gsc}). For each gold explanation set, a large language model is prompted to produce a storyboard over the set's resources; proposed role labels are canonicalised to the eleven-role vocabulary $\mathcal{R}$, yielding 741 reference storyboards, one per gold set (2.31 per query on average; mean 4.55 segments and 189\,s per storyboard). Because a query admits several valid explanation sets, a query with more than one gold set has multiple references, and a generated storyboard is scored against the best-matching one.

A sampled 219 storyboards undergo human validation by the same four annotators, each storyboard rated by two on six dimensions on an ordinal 0--2 scale, with overlapping assignments so that agreement can be measured on the same storyboards. Agreement (Table~\ref{tab:sb_iaa}) is high for grounding ($\alpha=0.70$), faithfulness/non-contradiction ($\alpha=0.71$), and on-screen fidelity ($\alpha=0.64$), and lower for narrative coherence ($\alpha=0.21$) and visual-generation readiness ($\alpha=0.29$); the lower coherence agreement indicates that pedagogical organisation is more subjective than grounding and resource fidelity, so reference-matching scores should be read alongside this uncertainty. Both layers are released.

\begin{table}[t]
\centering
\small
\begin{tabular}{lcc}
\toprule
Dimension & Krippendorff's $\alpha$ & Raw agreement \\
\midrule
Groundedness                     & 0.70 & 91\% \\
Faithfulness / non-contradiction & 0.71 & 90\% \\
On-screen fidelity               & 0.64 & 94\% \\
Narrative coherence              & 0.21 & 94\%   \\
Visual-generation readiness      & 0.29 & 94\%   \\
\midrule
Pooled                           & 0.62 & 93\% \\
\bottomrule
\end{tabular}
\caption{Inter-annotator agreement on the storyboard layer, over 219 sampled storyboards rated by four annotators on six ordinal dimensions. Agreement is high on grounding and faithfulness and lower on the more subjective coherence and readiness dimensions.}
\label{tab:sb_iaa}
\end{table}

\subsection{Final Benchmark}

EduExplain is built from secondary-level science textbooks. Extraction yields 6{,}921 typed resources---4{,}594 explanatory passages, 1{,}717 figures, 495 worked examples, and 115 activities---annotated with 2{,}123 concepts derived from chapter and section topics. The benchmark provides, for each of 321 learner-oriented queries (seven intents, three surface forms each), one or more human-validated gold explanation sets, and, as a companion layer, 741 human-reviewed reference storyboards. Table~\ref{tab:composition} summarizes the explanation-set composition by query category: no gold set exceeds three resources, and categories differ substantially in their number of alternative gold sets $m$---Example and Revision queries were each validated with three alternatives, whereas Visual, Misconception, and Practice queries were predominantly validated with one. Alongside the 321 validated queries, we release 992 \emph{silver} queries whose gold sets come from the same pipeline without the human-validation stage, disjoint from the validated set and useful for pre-training.

\section{Evaluation Metrics}

Unlike conventional retrieval benchmarks, EduExplain evaluates the retrieval of explanation sets rather than individual documents. Each query is associated with one or more human-validated explanation sets, and evaluation therefore considers both exact recovery of these sets and the conceptual completeness of the retrieved resources.

Throughout this section, let $G_q=\{G_q^{(1)},\ldots,G_q^{(m)}\}$ denote the human-validated explanation sets associated with query $q$, and let $R_q$ denote the retrieved explanation set.

We report two primary benchmark metrics—GoldSet F1 and Concept Coverage—and three auxiliary measures used for analysing particular experimental settings.

\subsection{GoldSet F1}

The primary objective of EduExplain is to retrieve one valid explanation set for a learner query. Since multiple human-validated explanation sets may exist, the retrieved set is compared against every gold explanation set and the best match is retained.

Gold set recall is defined as
\[
\text{GoldSetRecall}(q)=
\max_i
\frac{|R_q\cap G_q^{(i)}|}
{|G_q^{(i)}|},
\]
while precision is
\[
\text{GoldSetPrecision}(q)=
\max_i
\frac{|R_q\cap G_q^{(i)}|}
{|R_q|}.
\]
The primary benchmark metric is the corresponding F1 score,
\[
\text{GoldSetF1}(q)=
\max_i
\frac{2|R_q\cap G_q^{(i)}|}
{|R_q|+|G_q^{(i)}|}.
\]
Only exact resource matches are credited. Alternative pedagogically valid explanations are represented explicitly through multiple human-validated gold explanation sets rather than inferred through semantic similarity between resources.

\subsection{Core Concept Coverage}
\label{sec:concept_coverage}

Exact agreement with a gold explanation set does not fully characterize explanation quality. Different educational resources may explain the same concepts while using different examples, diagrams, or activities. We therefore additionally evaluate the conceptual completeness of the retrieved explanation.

Each resource is annotated with one or more educational concepts during benchmark construction. The concepts associated with a gold explanation set are obtained by taking the union of the concepts assigned to its constituent resources.

Given the collection of gold explanation sets for a query, the frequency of concept $c$ is
\[
\mathrm{freq}(c,q)=
\frac{
|\{G_q^{(i)}:c\in\mathrm{Concepts}(G_q^{(i)})\}|
}{m}.
\]
Concepts appearing in at least a fraction $\tau$ of the gold explanation sets are treated as the core concepts required to answer the query, $\mathrm{Core}(q)=\{c:\mathrm{freq}(c,q)\ge\tau\}$, where $\tau=0.8$ throughout this paper. The concepts covered by the retrieved explanation are $\mathrm{Concepts}(R_q)=\bigcup_{r\in R_q}\mathrm{Concepts}(r)$, and Concept Coverage is computed as
\[
\mathrm{CoreConceptCoverage}(q)=
\frac{
|\mathrm{Core}(q)\cap\mathrm{Concepts}(R_q)|
}
{|\mathrm{Core}(q)|}.
\]
Core Concept Coverage measures whether the retrieved explanation addresses the central concepts required by the query, regardless of which specific resources were selected. It complements GoldSet F1 by distinguishing failures caused by missing concepts from those arising from alternative but conceptually equivalent resource selections. The measure is defined only for queries with at least two gold explanation sets ($m\ge2$) and a non-empty core: when a query has a single gold set, every concept it contains trivially clears the threshold $\tau=0.8$ because there are no alternative sets to disagree, so core concepts are undefined in the intended sense and the query is excluded. Excluding $99$ single-gold-set queries and $2$ with an empty core leaves $220$ queries, which form the denominator wherever Core Concept Coverage is reported.

\subsection{Auxiliary Analysis Metrics}
\label{sec:aux_metrics}

The following measures are used in specific analyses but are not primary benchmark metrics.

\paragraph{Controlled GoldSet F1.}
GoldSet F1 increases with the number of annotated gold explanation sets because it selects the best matching explanation. For comparisons across query groups with different values of $m$, we therefore report
\[
\mathrm{CtrlF1}(q)=
\frac1m
\sum_{i=1}^{m}
\frac{
2|R_q\cap G_q^{(i)}|
}
{|R_q|+|G_q^{(i)}|},
\]
which averages F1 across all gold explanation sets rather than taking the maximum.

\paragraph{Set Concept Coverage.}
Core Concept Coverage (Section~\ref{sec:concept_coverage}) is undefined for queries with a single gold set, which excludes the categories dominated by such queries from any per-category coverage comparison. For by-category analysis we therefore report a companion measure, defined for every query. For a single gold set $G_q^{(i)}$, let $C_i(q)=|\mathrm{Concepts}(R_q)\cap\mathrm{Concepts}(G_q^{(i)})|/|\mathrm{Concepts}(G_q^{(i)})|$ be the fraction of that set's concepts recovered by the retrieval. Set Concept Coverage averages this over the query's gold sets,
\[
\mathrm{SetCC}(q)=\frac{1}{m}\sum_{i=1}^{m} C_i(q).
\]
Unlike Core Concept Coverage, it does not threshold concepts by cross-set frequency and is defined for all queries including those with $m=1$; it is used only in per-category breakdowns, over all 321 queries.

\paragraph{Attainment.}
When the explanation-set budget is fixed, GoldSet F1 cannot always reach 1.0 because retrieved and gold explanation sets may differ in size. Let
\[
\mathrm{MaxF1}(q)=
\max_i
\frac{
2\min(|R_q|,|G_q^{(i)}|)
}
{|R_q|+|G_q^{(i)}|}
\]
denote the maximum achievable F1 for the query given the retrieval budget. We define
\[
\mathrm{Attainment}(q)=
\frac{\mathrm{GoldSetF1}(q)}
{\mathrm{MaxF1}(q)},
\]
which normalizes GoldSet F1 by its query-specific upper bound. Attainment is reported only when all systems are evaluated with the same explanation-set budget.

\paragraph{Recall@$K$.}
To facilitate comparison with traditional retrieval benchmarks, we also report Recall@$K$. A resource is considered relevant if it appears in at least one human-validated explanation set, $\mathrm{Rel}(q)=\bigcup_i G_q^{(i)}$, giving
\[
\mathrm{Recall@}K=
\frac{
|\mathrm{Top}K(q)\cap\mathrm{Rel}(q)|
}
{|\mathrm{Rel}(q)|}.
\]
Recall@$K$ evaluates the quality of the candidate retrieval stage independently of the subsequent explanation-set construction process.

\subsection{Storyboard Metrics}
\label{sec:sb_metrics}

The storyboard layer is evaluated against the reference storyboards along four axes; each metric takes the best match over a query's references. Let $\pi(B)=(\rho_1,\ldots,\rho_L)$ denote the role sequence of a storyboard $B$ and $\mathrm{res}(B)=\bigcup_\ell S_\ell$ its grounding resources.

\paragraph{Structure.}
\emph{Role-set F1} (RS-F1) is the multiset F1 between predicted and reference roles, $\max_{B^\star} 2PR/(P+R)$, assessing role composition independent of order. \emph{Role-sequence similarity} (RSS) additionally penalises ordering differences via normalised edit distance, $\max_{B^\star}[1-\mathrm{lev}(\pi(B),\pi(B^\star))/\max(|\pi(B)|,|\pi(B^\star)|)]$.

\paragraph{Grounding.}
\emph{Grounding faithfulness} (GF) is the fraction of segments whose grounding set is non-empty and contained in the input set, $L^{-1}\sum_\ell \mathbf{1}[\emptyset\neq S_\ell\subseteq R_q]$, measuring whether segments point to resources actually in the explanation set.

\paragraph{Coverage.}
\emph{Grounding coverage} (GCov) is $|\mathrm{res}(B)\cap R_q|/|R_q|$, the fraction of the available set the storyboard uses.

\paragraph{Timing.}
\emph{Budget adherence} is $\mathbf{1}[\,|T(B)-T^{\mathrm{tgt}}|\le\delta\,]$, whether the planned total duration meets the target within tolerance.

These measures are complementary: GF checks resource membership, GCov checks how much of the set is used, and RS-F1/RSS assess whether the plan's pedagogical role structure matches the references. Structural grounding and full coverage are therefore not mistaken for a well-organised explanation.

\section{Baselines}

EduExplain evaluates retrieval of explanation sets rather than individual documents or passages. Existing retrieval models rank individual resources, whereas the benchmark requires retrieving a compact collection of complementary educational resources that together support an explanation. We therefore evaluate two independent components of the retrieval pipeline: (i) candidate retrieval and (ii) explanation-set construction.

All baselines follow the same two-stage architecture. Given a learner query $q$, a retrieval model first produces a ranked candidate pool $C_q=\{r_1,\ldots,r_K\}$, with $K$ typically $100$, and an explanation-set selection algorithm then chooses a subset $R_q\subseteq C_q$ of at most $n$ resources. Combining multiple retrieval models with multiple selection strategies lets the benchmark separate improvements due to candidate retrieval from those due to explanation-set construction.

\subsection{Candidate Retrieval}

We evaluate representative lexical, dense, multimodal, and reranking approaches.

A fixed dense encoder is a configuration choice rather than part of the benchmark definition; the reference baselines below use BGE-M3 as the dense encoder, while any single-vector encoder may be substituted.

\paragraph{BM25.}
BM25 serves as the lexical baseline. Every resource is indexed using its textual representation: textual resources use their generated summaries, while figures are indexed using automatically generated descriptions. Retrieval is performed using standard BM25 scoring.

\paragraph{Dense Retrieval.}
For dense semantic retrieval, resources are represented by their textual summaries or descriptions and encoded using BGE-M3 (\texttt{bge-m3}). Queries and resources are ranked by cosine similarity between their embeddings.

\paragraph{Multimodal Retrieval.}
To evaluate retrieval using native visual information, we employ OpenCLIP (\texttt{ViT-L-14}). Textual resources are encoded using the text encoder, while figures are encoded directly using the image encoder. Queries are encoded using the text encoder and matched within the shared embedding space.

\paragraph{Cross-Encoder Reranking.}
Finally, we evaluate a two-stage pipeline in which the candidate pool produced by dense retrieval is reranked using Qwen3-VL-Reranker-2B. Since reranking only reorders an existing candidate pool, Recall@$K$ remains identical to that of the underlying dense retriever.

\subsection{Explanation-Set Construction}

Given the ranked candidate pool $C_q$, the second stage constructs an explanation set $R_q\subseteq C_q$ with $|R_q|\le n$, where $n$ is the explanation-set budget. We evaluate four selection strategies.

\paragraph{Top-$n$.}
The simplest baseline returns the highest-ranked resources, $R_q=\{r_1,\ldots,r_n\}$. It optimizes only retrieval relevance and does not account for redundancy or modality balance.

\paragraph{Modality-Capped Selection.}

Educational explanations often require both textual and visual resources. To encourage balance, we partition the candidate pool by modality, $C_q=\bigcup_{m\in\mathcal M} C_q^{(m)}$ with $\mathcal M=\{\text{text},\text{image}\}$, and set a maximum quota $\kappa_m$ per modality satisfying $\sum_{m}\kappa_m=n$. Scanning candidates in ranked order, a resource $r$ is added iff $|R_q^{(m(r))}|<\kappa_{m(r)}$, where $m(r)$ is the modality of $r$; selection terminates once $|R_q| = n$. We set $\kappa_{\text{text}}=2$ and $\kappa_{\text{image}}=1$ (summing to $n=3$). This preserves the retrieval ranking while preventing a single modality from dominating.

\paragraph{Maximum Marginal Relevance (MMR).}

MMR iteratively selects
\[
r^*
=
\arg\max_{r\in C_q\setminus R_q}
\left[
\lambda\,\mathrm{Rel}(q,r)
-
(1-\lambda)
\max_{r'\in R_q}
\mathrm{Sim}(r,r')
\right],
\]
where $\mathrm{Rel}(q,r)$ is the retrieval score, $\mathrm{Sim}(r,r')$ is computed from BGE-M3 embeddings of the resource summaries, and $\lambda=0.8$ trades relevance against diversity; the process repeats until $|R_q|=n$.

\paragraph{Modality-Aware MMR.}

We extend MMR by penalizing modalities already well represented in the set. With $N_m(R_q)=|\{r\in R_q:m(r)=m\}|$ the number of selected resources from modality $m$, the selection score becomes
\[
\begin{aligned}
\mathrm{Score}(r)
&=
\lambda\,\mathrm{Rel}(q,r)
-(1-\lambda)
\max_{r'\in R_q}\mathrm{Sim}(r,r')\\
&\quad
-\gamma
\frac{N_{m(r)}(R_q)}
{\kappa_{m(r)}},
\end{aligned}
\]
where $\gamma$ controls the penalty strength; we use $\lambda=0.8$ and $\gamma=0.1$, with the same modality quotas as above. The candidate with the highest score is added at each iteration, encouraging explanations that combine complementary textual and visual resources while retaining the diversity benefits of MMR.

\subsection{Experimental Configurations}

Every retrieval model is evaluated with every explanation-set construction strategy, producing the sixteen retrieval pipelines summarized in Table~\ref{tab:rq1}. We analyze the improvements due to candidate retrieval and those due to explanation-set construction, allowing future retrieval or selection methods to be incorporated independently.

\subsection{Storyboard-Construction Baselines}
\label{sec:sb_baselines}

For the structuring stage we provide two reference baselines that take a query and gold explanation set as input and produce a storyboard, isolating two obvious failure modes. \textsc{FreeForm} prompts a language model to generate a whole storyboard with no structural constraints or grounding checks. \textsc{FixedContract} imposes a fixed role progression (\textsc{hook}, \textsc{definition}, \textsc{mechanism}, \textsc{summary}) and fills each slot from the input set. Table~\ref{tab:sb_baselines} reports both against the reference storyboards. \textsc{FreeForm} achieves low resource-membership grounding (GF 0.595) and rarely meets the time budget (13.9\%), because an unconstrained generator drifts outside the input set and ignores duration; \textsc{FixedContract} satisfies grounding and budget by construction but recovers the reference role composition only moderately (RS-F1 0.713). Neither reaches the reference plans, confirming that storyboard construction is a non-trivial task the benchmark can measure: unconstrained generation fails grounding and timing, while a rigid template fails to adapt its role structure to the query.

\begin{table}[t]
\centering
\small
\begin{tabular}{lccccc}
\toprule
Method & RS-F1 & RSS & GF & GCov & Budget\% \\
\midrule
\textsc{FreeForm}      & 0.607 & 0.441 & 0.595 & 0.959 & 13.9 \\
\textsc{FixedContract} & 0.713 & 0.666 & 1.000 & 0.796 & 100.0 \\
\bottomrule
\end{tabular}
\caption{Storyboard-construction reference baselines, evaluated against the reference storyboards. RS-F1 and RSS assess role composition and sequence; GF and GCov assess grounding and coverage; Budget\% is the fraction of plans meeting the time budget. Grounding and budget are enforced by construction for \textsc{FixedContract}.}
\label{tab:sb_baselines}
\end{table}

\section{Experimental Evaluation}
\label{sec:experimental_evaluation}

The objective of the experimental evaluation is to understand the contribution of the different components of an explanation-set retrieval pipeline. To isolate the effect of each component, every experiment varies only a single independent variable while keeping all remaining components fixed.

Unless otherwise stated, experiments use the default pipeline identified in RQ1: dense retrieval with cross-encoder reranking and Top-$n$ selection, with candidate-pool size $K{=}100$ and explanation-set budget $n{=}3$. Each experiment then varies a single component of this pipeline and holds the rest fixed. Except in RQ7, which studies phrasing directly, each of the 321 information needs enters with one of its three surface forms chosen uniformly at random, so the reported numbers reflect a representative mix of learner phrasings rather than a single best-case wording.

\subsection{Significance Testing}
\label{sec:significance}

Three comparisons single out one pipeline component each -- the reranker (RQ1), the budget $n$ (RQ3), and the selection strategy (RQ4) -- tested with the paired protocol of Table~\ref{tab:significance}. Because every system runs on the same $N{=}321$ queries, we compare query by query rather than on means, which removes query-to-query variance in difficulty and is far more powerful than comparing averages; since each contrast varies one component and fixes the rest, a significant difference is attributable to that component alone. We assess significance with the two-sided Wilcoxon signed-rank test on per-query GoldSet F1 differences, reporting the statistics defined in Table~\ref{tab:significance}. RQ1 is a single pre-registered contrast, reported unadjusted; RQ3's two and RQ4's three contrasts each form a Holm--Bonferroni-corrected family. Where a null result is central -- notably Top-$n$ versus MMR (RQ4) -- we add a two one-sided equivalence test (TOST) so that absence of evidence is not misread as equivalence. Per-contrast findings appear in the corresponding RQ sections.

\subsection{RQ1: Retrieval–selection pipelines}
\label{sec:rq1}

\begin{table*}[t]
\centering
\small
\setlength{\tabcolsep}{4pt}
\resizebox{\textwidth}{!}{%
\begin{tabular}{l|c|ccc|ccc|ccc|ccc}
\toprule
\multirow{3}{*}{Retriever}
& \multirow{3}{*}{Recall@$K$}
& \multicolumn{12}{c}{Explanation-set selection method} \\
\cmidrule(lr){3-14}
& & \multicolumn{3}{c}{Top-$n$}
  & \multicolumn{3}{c}{Modality-Capped}
  & \multicolumn{3}{c}{MMR}
  & \multicolumn{3}{c}{Modality-Aware MMR} \\
\cmidrule(lr){3-5}\cmidrule(lr){6-8}\cmidrule(lr){9-11}\cmidrule(lr){12-14}
& & GS-F1 & Att & Core CC
  & GS-F1 & Att & Core CC
  & GS-F1 & Att & Core CC
  & GS-F1 & Att & Core CC \\
\midrule
BM25
& 0.830
& \textbf{0.349} & 0.399 & 0.678
& 0.330 & 0.377 & 0.686
& 0.348 & 0.397 & 0.678
& 0.349 & 0.398 & 0.681 \\
Dense retrieval (BGE-M3)
& 0.876
& \textbf{0.421} & 0.486 & 0.747
& 0.380 & 0.430 & 0.742
& 0.363 & 0.420 & 0.717
& 0.337 & 0.386 & 0.709 \\
Multimodal retrieval (OpenCLIP)
& 0.546
& \textbf{0.245} & 0.275 & 0.609
& 0.242 & 0.277 & 0.539
& 0.194 & 0.219 & 0.536
& 0.194 & 0.219 & 0.536\\
Dense $+$ cross-encoder reranking
& 0.876
& \underline{\textbf{0.461}} & 0.531 & 0.810
& 0.419 & 0.481 & 0.794 & 0.392 & 0.455 & 0.751 & 0.367 & 0.427 & 0.728 \\
\bottomrule
\end{tabular}%
}
\caption{
GoldSet F1 (GS-F1), attainment (Att), and Core Concept Coverage (Core CC) for every retriever (rows) and explanation-set selection method (columns), at $K = 100$ and $n = 3$. Recall@K evaluates the candidate pool and is reported once per retriever. GS-F1 and attainment are computed over all 321 queries; Core Concept Coverage over the 220 queries where it is defined. Bold marks the best GS-F1 per retriever; underline marks the best overall.}
\label{tab:rq1}
\end{table*}

Table~\ref{tab:rq1} reports both candidate-pool quality (Recall@$K$) and final-set quality (GS-F1, attainment, and Core Concept Coverage) for each of the sixteen retriever--selector pipelines, at $K=100$ and $n=3$. Two patterns emerge. First, first-stage retrieval quality dominates: dense retrieval (BGE-M3) reaches Recall@$K=0.876$ against $0.830$ for BM25 and only $0.546$ for OpenCLIP, and this ordering carries through to final-set quality. The multimodal retriever underperforms substantially (GS-F1 $\approx 0.245$), indicating that native CLIP-style image embeddings are weaker than textual summaries of the same diagrams for recovering the annotator-selected resources. Second, cross-encoder reranking of the dense pool yields the largest single improvement, raising GS-F1 from $0.421$ to $0.461$ and Core CC from $0.747$ to $0.810$ while leaving Recall@$K$ unchanged by construction; this $+0.04$ GS-F1 gain is reliable (Table~\ref{tab:significance}, $p=4.7{\times}10^{-3}$). The strongest complete pipeline is therefore dense retrieval with cross-encoder reranking paired with Top-$n$ selection (GS-F1 $=0.461$), which we adopt as the default configuration in all subsequent experiments. GoldSet F1 is the primary selection criterion, since exact explanation-set recovery is the principal benchmark objective, with Core CC breaking ties between configurations of comparable GS-F1. Following Section~\ref{sec:concept_coverage}, Core CC is reported over the $220$ queries with at least two gold sets, for which core concepts are well defined.

\begin{table*}[t]
\centering
\small
\setlength{\tabcolsep}{4pt}
\resizebox{\textwidth}{!}{%
\begin{tabular}{llcccccc}
\hline
RQ&Comparison&$\Delta$&95\% CI&$W$&$p$&$r$&Win/Tie/Loss\\
\hline
RQ1&w/ Reranker vs.\ w/o Reranker&$+0.040$&$[+0.011, +0.069]$&13764.0&{\boldmath$4.7{\times}10^{-3}$}$^{\dagger}$&$+0.23$&89/176/56\\
\hline
RQ3&$n{=}3$ vs.\ $n{=}2$&$+0.004$&$[-0.016,+0.023]$&22418.5&$0.1$&$-0.10$&81/53/187\\
RQ3&$n{=}3$ vs.\ $n{=}4$&$+0.007$&$[-0.009, +0.022]$&19968.0&{\boldmath$1.8{\times}10^{-3}$}&$+0.20$&213/37/71\\
\hline
RQ4&Top-$n$ vs.\ MMR&$+0.069$&$[+0.046, +0.093]$&6409.5&$\mathbf{7.4\times 10^{-8}}$&$+0.54$&80/217/24\\
RQ4&Top-$n$ vs.\ Modality-Capped&$+0.041$&$[+0.021, +0.062]$&7426.0&{\boldmath$1.9{\times}10^{-4}$}&$+0.40$&62/232/27\\
RQ4&Top-$n$ vs.\ Modality-Aware MMR&$+0.094$&$[+0.070, +0.119]$&5997.0&{\boldmath$1.3{\times}10^{-11}$}&$+0.62$&100/197/24\\
\hline
\end{tabular}%
}
\caption{Paired significance tests for the three central comparisons, on per-query GoldSet F1 over all $N{=}321$ queries. Each row varies one component of the default pipeline (dense retrieval with cross-encoder reranking, Top-$n$, $K{=}100$, $n{=}3$): RQ1 removes the reranker, RQ3 changes the budget $n$, RQ4 changes the selection strategy. $\Delta$ is the mean paired GS-F1 difference (positive favours the first-named system) with a 95\% paired-bootstrap CI ($10,000$ resamples of queries, fixed seed); $W$, $p$ are the two-sided Wilcoxon signed-rank statistic and its $p$-value; $r$ is the rank-biserial effect size; Win/Tie/Loss counts queries. RQ1 is reported unadjusted ($^{\dagger}$); the RQ3 and RQ4 contrasts are Holm--Bonferroni corrected within their families. Bold marks $p<0.05$ after correction.}
\label{tab:significance}
\end{table*}

\subsection{RQ2: Candidate Pool Size}

Table \ref{tab:rq2} varies the candidate-pool size $K$ while holding the retriever, selector, and explanation-set budget fixed. As $K$ grows from 25 to 200, $Recall@K$ rises steadily from 0.764 to 0.921, confirming that larger pools do surface more of the gold resources. Final-set quality, however, is essentially flat: GS-F1 stays within 0.458–0.464 (peaking at $K = 100$ and $K = 50$) and Core Concept Coverage within 0.810–0.815. The gap between the rising $Recall@K$ and the stationary GS-F1 shows that the binding constraint is the explanation-set budget $n$, not the size of the candidate pool: once the relevant resources are present in the pool, enlarging it further only adds distractors that the fixed-size selection stage cannot exploit. We therefore fix $K = 100$ in the remaining experiments.

\begin{table}[t]
\centering
\small
\begin{tabular}{cccc}
\hline
Candidate Pool ($K$) & GS-F1 & Core CC & Recall@$K$\\
\hline
25 & 0.458 & 0.812 & 0.764\\
50 & 0.461 & 0.814 & 0.832\\
100 & 0.461 & 0.810 & 0.876\\
200 & 0.464 & 0.815 & 0.921\\
\hline
\end{tabular}
\caption{Effect of candidate-pool size $K$ on final-set quality and candidate recall, for the default pipeline (with $n{=}3$). GS-F1 and Concept Coverage evaluate the final explanation set; Recall@$K$ evaluates the candidate pool. Core Concept Coverage is over the 220 queries where it is defined.}
\label{tab:rq2}
\end{table}

\subsection{RQ3: Explanation-Set Size}

Table \ref{tab:rq3} varies the explanation-set budget $n$ from 2 to 6. Precision falls and recall rises monotonically with $n$, as expected: larger sets recover more gold resources but admit more non-gold ones. GoldSet F1, which balances the two, peaks at $n = 3 (0.461)$. This optimum coincides with the budget that admits the highest achievable score: because no gold explanation set in the benchmark contains more than three resources, the mean upper bound Max. F1 is maximised at $n = 3$ (0.892) and drops sharply for $n \geq 4$, where the retrieved set is necessarily larger than any gold set and F1 can no longer reach 1.0. Concretely, $n = 3$ gives a perfect score for 179 of the 321 queries, whereas $n \geq 4$ does not do so for any query. A Friedman omnibus over $n\in\{2,\ldots,6\}$ confirms that budget significantly affects GoldSet F1 ($\chi^2(4){=}114.7$, $p<10^{-20}$); in the paired tests (Table~\ref{tab:significance}), $n{=}3$ significantly exceeds $n{=}4$ but not $n{=}2$, so the choice of $n{=}3$ rests on the ceiling argument above rather than on a pairwise win over $n{=}2$. Core CC continues to rise with $n$ (from 0.731 at $n = 2$ to 0.902 at $n = 6$), since larger sets touch more concepts, but this comes at the cost of precision and does not reflect better explanation-set recovery. We therefore fix $n = 3$ as the default budget: it maximises both GoldSet F1 and the achievable ceiling, and Core CC past this point reflects set size rather than retrieval quality.

\begin{table}[t]
\centering
\small
\begin{tabular}{cccccc}
\hline
Set Size ($n$)&GS-Precision&GS-Recall&GS-F1& Core CC &Max.\ F1\\
\hline
2&\textbf{0.497}&0.449&0.457&0.731&0.878\\
3&0.416&0.548&\textbf{0.461}&0.810&\textbf{0.892}\\
4&0.368&0.629&0.454&0.853&0.755\\
5&0.320&0.681&0.427&0.876&0.655\\
6&0.287&\textbf{0.723}&0.404&\textbf{0.902}&0.579\\
\hline
\end{tabular}
\caption{Effect of the explanation-set budget $n$ on precision, recall, GoldSet F1, and Core Concept Coverage, for the default pipeline. Max. F1 is the mean per-query upper bound on GS-F1 imposed by the budget. Core Concept Coverage is computed over the 220 queries where it is defined.}
\label{tab:rq3}
\end{table}

\subsection{RQ4: Explanation-Set Selection}
\label{sec:rq4}

Table \ref{tab:rq4} fixes the first-stage retriever to the reranked dense pipeline and varies only the selection strategy. On this strong ranking, Top-$n$ selection achieves the highest GoldSet F1 (0.461), narrowly ahead of Modality-Capped Selection (0.419), with MMR (0.392) and Modality-Aware MMR (0.367) trailing. That all three set-aware strategies fall below naive Top-n is the central finding of this experiment, and it points to a fundamental mismatch between semantic diversity and explanatory sufficiency. The two modality-constrained selectors are significantly worse than Top-$n$ (Table~\ref{tab:significance}). The Top-$n$--MMR gap, by contrast, is not significant: the two strategies return identical sets on 285 of 321 queries ($88.8\%$), a sign test on the 36 differing pairs is non-significant (23 wins vs.\ 13 losses, $p{=}0.13$), and a pre-specified equivalence test (TOST, $\delta{=}0.02$, fixed in advance as roughly twice the observed Top-$n$--MMR gap) is also inconclusive (90\% bootstrap CI on $\Delta$: $[+0.000,+0.022]$, marginally outside the margin). We therefore read this as absence of evidence that diversity helps, not as established equivalence.
The interaction with the first-stage ranking confirms that the mismatch is not an artefact of reranking. Top-$n$ already leads MMR on the plain dense retriever (Table \ref{tab:rq1}, $0.421$ vs.\ $0.363$), and cross-encoder reranking only widens the gap ($0.461$ vs.\ $0.392$): scoring each candidate against the query directly concentrates the gold resources in the highest ranks, so there is even less for a diversity penalty to usefully remove.

In this setting MMR's diversity penalty misfires: the resources that make up a good explanation are often semantically close -- a definition, an explanatory passage, and a worked example of the same concept overlap heavily in content -- so MMR penalizes exactly the resources it should keep, and the modality-constrained variants fail for the same reason, optimizing a proxy (embedding dissimilarity, modality balance) that is not aligned with explanatory sufficiency. \emph{Explanation-set construction is therefore not just a diversification problem}: a good set is complementary in function (define, illustrate, visualize, assess), and this need not correspond to embedding distance. Off-the-shelf diversity objectives thus have no reliable signal to exploit and, on a well-ranked pool, the two modality-constrained variants do more harm than good. Closing the gap between the reranked Top-$n$ baseline and a genuinely explanation-aware selector will require modeling the roles resources play and the sufficiency of the set as a whole rather than pairwise redundancy -- one of the main open problems the benchmark surfaces. Consistent with this, Core CC is nearly flat across strategies (0.778--0.798): the strategies differ in which resources they select, not in the concepts covered, so the losses come from discarding valid explanatory resources rather than missing conceptual content.

\begin{table}[t]
\centering
\small
\begin{tabular}{lccccc}
\hline
Selection Strategy&GS-Prec &GS-Recall&GS-F1&Att& Core CC \\
\hline
Top-$n$&\textbf{0.416}&\textbf{0.548}&\textbf{0.461}&\textbf{0.531}&\textbf{0.810}\\
Modality-Capped&0.380&0.498&0.419&0.481&0.794\\
MMR&0.352&0.468&0.392&0.455&0.751\\
Modality-Aware MMR&0.329&0.440&0.367&0.427&0.729\\
\hline
\end{tabular}
\caption{Comparison of the four explanation-set selection strategies on the reranked dense retriever ($K = 100, n = 3$), reporting precision, recall, GoldSet F1, attainment, and Core Concept Coverage. Core Concept Coverage is computed over the 220 queries where it is defined.}
\label{tab:rq4}
\end{table}

\subsection{RQ5: Performance Across Query Types}
\label{sec:rq5}

Table \ref{tab:rq5} breaks performance down by the seven query categories. Because GoldSet F1 is maximised over a query's gold sets, categories with more gold sets (higher $m$) enjoy an inflated raw F1; we therefore read category differences off Ctrl. F1, which averages over gold sets and is invariant to $m$. Under this controlled measure, Example (0.382) and Explanation (0.423) are the strongest categories and Misconception (0.246) is by far the weakest, with Visual (0.339) in the middle rather than at the bottom. Two categories deserve comment. Misconception queries are hard under every measure: they are phrased as plausible-but-wrong beliefs (e.g. "Do plants get food from the soil?"), so lexical and semantic overlap with the correcting resources is low, and the pipeline struggles to surface the resource that addresses the misconception rather than the surface topic. Visual queries show the opposite profile -- a low GoldSet F1 (0.348) but the highest Set Concept Coverage of any category (0.765). This gap is diagnostic: the retrieved resources cover the concepts the query requires but are not the specific diagrams the annotators selected, so the pipeline is finding conceptually correct material while missing the particular visual resource. This is consistent with the weak standalone performance of image retrieval seen in RQ1 (Table~\ref{tab:rq1}) and suggests that visual explanation is a category where better image-side retrieval, rather than better set selection, is the bottleneck.

\begin{table}[t]
\centering
\small
\begin{tabular}{lcccccc}
\hline
Query Type&$N$&$m$&GS-F1&Att&Ctrl.\ F1& Set CC\\
\hline
Example&66&3.00&0.517&0.577&0.382&0.711\\
Explanation&32&2.59&\textbf{0.554}&\textbf{0.640}&\textbf{0.423}&0.744\\
Why/How&54&2.80&0.500&0.553&0.399&0.709\\
Revision&61&3.00&0.454&0.497&0.335&0.642\\
Practice&57&1.18&0.444&0.476&0.355&0.625\\
Visual&27&1.07&0.348&0.574&0.339&\textbf{0.765}\\
Misconception&24&1.25&0.275&0.375&0.246&0.594\\
\hline
\end{tabular}
\caption{Performance by query category, for the default pipeline. $N$ is the number of queries in the category and m their mean number of gold explanation sets. Because GS-F1 is maximised over a query's m gold sets, Ctrl. F1 is reported as an m-invariant control that averages over gold sets instead of taking the maximum. Set CC is Set Concept Coverage (Section~\ref{sec:aux_metrics}), defined for all 321 queries, unlike the Core Concept Coverage of the other tables.}
\label{tab:rq5}
\end{table}

\subsection{RQ6: Query Complexity}

Table \ref{tab:rq6} groups queries by the number of concepts they are generated from. Performance is non-monotonic in complexity: two-concept queries score lowest on every measure (Ctrl. F1 = 0.269, Set CC = 0.590), below both single-concept (0.481, 0.698) and three-concept (0.499, 0.705) queries. This dip is compositional, not intrinsic -- the hardest RQ5 categories are concentrated in the two-concept bin (33\% are Misconception or Visual, against 11\% and 14\% for one- and three-concept queries). Once category is controlled for, concept count is not by itself a strong predictor of difficulty on this benchmark: difficulty is driven by the pedagogical intent of a query rather than its conceptual breadth.

\begin{table}[t]
\centering
\small
\begin{tabular}{lccccc}
\hline
Number of Concepts&$N$&$m$&GS-F1&Ctrl.\ F1& Set CC\\
\hline
1 (single)&160&2.58&0.481&	0.379&0.698\\
2 (pair)&57&1.61&0.334&0.269&0.590\\
3 (triple)&104&2.28&\textbf{0.499}&\textbf{0.385}&\textbf{0.705}\\
\hline
\end{tabular}
\caption{Performance by query complexity, measured as the number of concepts a query is generated from, for the default pipeline. N is the number of queries and m their mean number of gold explanation sets.}
\label{tab:rq6}
\end{table}

\subsection{RQ7: Robustness to Query Formulation}

Table~\ref{tab:rq7} groups the 321 queries by the surface form that was drawn for them, giving a first read on whether phrasing affects retrieval. The spread is small: GoldSet F1 varies by only 0.06 across the three variants and Ctrl. F1 by 0.039, and the ordering of Search and Chat reverses between the two measures, placing their difference within noise. The one consistent signal is that learning-context formulations score highest on every measure (GS-F1 = 0.497, Ctrl. F1 = 0.387, Set CC = 0.705), which is plausible: a learning-oriented question carries more explicit topical and pedagogical context than a terser query, giving the retriever and reranker more to match against. This comparison is suggestive rather than controlled: because each need contributes only its randomly drawn variant, the three rows contain different needs, so the gap confounds phrasing with which needs happened to fall in each group (the rows also differ slightly in mean gold-set count $m$). A fully controlled test would hold the need fixed and compare its three variants paired by parent query; we leave this paired analysis to future work.

\begin{table}[t]
\centering
\small
\begin{tabular}{lccccc}
\hline
Query Variant&$N$&$m$&GS-F1&Ctrl.\ F1&Set CC\\
\hline
Search&110&2.28&0.437&0.349&0.668\\
Chat&103&2.36&0.448&0.348&0.670\\
Learning Context&108&2.29&\textbf{0.497}&\textbf{0.387}&\textbf{0.705}\\
\hline
\end{tabular}
\caption{Retrieval robustness across query formulations, for the default pipeline. Each learner information need is expressed three ways — a search-engine query, a conversational turn, and a learning-oriented question. $N$ is the number of queries and $m$ their mean number of gold explanation sets}
\label{tab:rq7}
\end{table}

\section{Conclusion and Research Directions}
\label{sec:conclusion}

We presented EduExplain, a benchmark that evaluates the two stages of educational explanation generation that precede rendering: retrieving a compact set of complementary resources, and structuring that set into an ordered, timed, role-assigned plan. EduExplain provides human-validated ground truth for both---321 queries with explanation sets over 6{,}921 typed multimodal resources, and 741 reference storyboards---together with stage-appropriate metrics and reference baselines. Our baselines show that retrieval-stage quality is governed by the first-stage retriever, with cross-encoder reranking the single largest gain, and that off-the-shelf diversity-based selection underperforms simple top-ranked selection because the resources of a good explanation are often semantically close; on the structuring stage, both unconstrained and rigidly templated construction fall short of the reference plans.

These results point to several open problems the benchmark surfaces. \emph{Set-sufficiency modeling}: rather than scoring items for pairwise redundancy, a selector should score a whole set for explanatory completeness. \emph{Pedagogical image retrieval}: visual queries attain high concept coverage but low exact recovery, so the pipeline finds conceptually correct material yet misses the specific figure, isolating image-side retrieval of near-duplicate figures as a bottleneck. \emph{Misconception-aware retrieval}: the hardest category requires retrieving the resource that corrects a plausible-but-wrong belief, which has low lexical and semantic overlap with the query. \emph{Explanation-aware structuring}: closing the gap between templated construction and the reference storyboards requires modeling pedagogical role structure rather than fixed progressions. By providing stage-wise ground truth, EduExplain offers a testbed for all four.

\bibliographystyle{plain}
\bibliography{mmir}

@inproceedings{fever,
  author     = {James Thorne and
Andreas Vlachos and
Christos Christodoulopoulos and
Arpit Mittal},
  bibsource  = {dblp computer science bibliography, https://dblp.org},
  booktitle  = {Proceedings of the 2018 Conference of the North American Chapter of
the Association for Computational Linguistics: Human Language Technologies,
{NAACL-HLT} 2018, New Orleans, Louisiana, USA, June 1-6, 2018, Volume
1 (Long Papers)},
  doi        = {10.18653/V1/N18-1074},
  editor     = {Marilyn A. Walker and
Heng Ji and
Amanda Stent},
  pages      = {809--819},
  publisher  = {Association for Computational Linguistics},
  title      = {{FEVER:} a Large-scale Dataset for Fact Extraction and VERification},
  url        = {https://doi.org/10.18653/v1/n18-1074},
  year       = {2018}
}

@inproceedings{scifact,
  author     = {David Wadden and
Shanchuan Lin and
Kyle Lo and
Lucy Lu Wang and
Madeleine van Zuylen and
Arman Cohan and
Hannaneh Hajishirzi},
  bibsource  = {dblp computer science bibliography, https://dblp.org},
  booktitle  = {Proceedings of the 2020 Conference on Empirical Methods in Natural
Language Processing, {EMNLP} 2020, Online, November 16-20, 2020},
  doi        = {10.18653/V1/2020.EMNLP-MAIN.609},
  editor     = {Bonnie Webber and
Trevor Cohn and
Yulan He and
Yang Liu},
  pages      = {7534--7550},
  publisher  = {Association for Computational Linguistics},
  title      = {Fact or Fiction: Verifying Scientific Claims},
  url        = {https://doi.org/10.18653/v1/2020.emnlp-main.609},
  year       = {2020}
}

@inproceedings{visdom,
  author     = {Manan Suri and
Puneet Mathur and
Franck Dernoncourt and
Kanika Goswami and
Ryan A. Rossi and
Dinesh Manocha},
  bibsource  = {dblp computer science bibliography, https://dblp.org},
  booktitle  = {Proceedings of the 2025 Conference of the Nations of the Americas
Chapter of the Association for Computational Linguistics: Human Language
Technologies, {NAACL} 2025 - Volume 1: Long Papers, Albuquerque, New
Mexico, USA, April 29 - May 4, 2025},
  doi        = {10.18653/V1/2025.NAACL-LONG.310},
  editor     = {Luis Chiruzzo and
Alan Ritter and
Lu Wang},
  pages      = {6088--6109},
  publisher  = {Association for Computational Linguistics},
  title      = {VisDoM: Multi-Document {QA} with Visually Rich Elements Using Multimodal
Retrieval-Augmented Generation},
  url        = {https://doi.org/10.18653/v1/2025.naacl-long.310},
  year       = {2025}
}

@article{mmlongbench,
  author     = {Zhaowei Wang and
Wenhao Yu and
Xiyu Ren and
Jipeng Zhang and
Yu Zhao and
Rohit Saxena and
Liang Cheng and
Ginny Y. Wong and
Simon See and
Pasquale Minervini and
Yangqiu Song and
Mark Steedman},
  bibsource  = {dblp computer science bibliography, https://dblp.org},
  doi        = {10.48550/ARXIV.2505.10610},
  eprint     = {2505.10610},
  eprinttype = {arXiv},
  journal    = {CoRR},
  title      = {MMLongBench: Benchmarking Long-Context Vision-Language Models Effectively
and Thoroughly},
  url        = {https://doi.org/10.48550/arXiv.2505.10610},
  volume     = {abs/2505.10610},
  year       = {2025}
}

@inproceedings{mmdocir,
  author     = {Kuicai Dong and
Yujing Chang and
Derrick{-}Goh{-}Xin Deik and
Dexun Li and
Ruiming Tang and
Yong Liu},
  bibsource  = {dblp computer science bibliography, https://dblp.org},
  booktitle  = {Proceedings of the 2025 Conference on Empirical Methods in Natural
Language Processing, {EMNLP} 2025, Suzhou, China, November 4-9, 2025},
  doi        = {10.18653/V1/2025.EMNLP-MAIN.1576},
  editor     = {Christos Christodoulopoulos and
Tanmoy Chakraborty and
Carolyn Rose and
Violet Peng},
  pages      = {30971--31005},
  publisher  = {Association for Computational Linguistics},
  title      = {MMDocIR: Benchmarking Multimodal Retrieval for Long Documents},
  url        = {https://doi.org/10.18653/v1/2025.emnlp-main.1576},
  year       = {2025}
}

@inproceedings{mmr,
  author     = {Jaime G. Carbonell and
Jade Goldstein},
  bibsource  = {dblp computer science bibliography, https://dblp.org},
  booktitle  = {{SIGIR} '98: Proceedings of the 21st Annual International {ACM} {SIGIR}
Conference on Research and Development in Information Retrieval, August
24-28 1998, Melbourne, Australia},
  doi        = {10.1145/290941.291025},
  editor     = {W. Bruce Croft and
Alistair Moffat and
C. J. van Rijsbergen and
Ross Wilkinson and
Justin Zobel},
  pages      = {335--336},
  publisher  = {{ACM}},
  title      = {The Use of MMR, Diversity-Based Reranking for Reordering Documents
and Producing Summaries},
  url        = {https://doi.org/10.1145/290941.291025},
  year       = {1998}
}

@inproceedings{kembhavi,
  author     = {Aniruddha Kembhavi and
Min Joon Seo and
Dustin Schwenk and
Jonghyun Choi and
Ali Farhadi and
Hannaneh Hajishirzi},
  bibsource  = {dblp computer science bibliography, https://dblp.org},
  booktitle  = {2017 {IEEE} Conference on Computer Vision and Pattern Recognition,
{CVPR} 2017, Honolulu, HI, USA, July 21-26, 2017},
  doi        = {10.1109/CVPR.2017.571},
  pages      = {5376--5384},
  publisher  = {{IEEE} Computer Society},
  title      = {Are You Smarter Than a Sixth Grader? Textbook Question Answering for
Multimodal Machine Comprehension},
  url        = {https://doi.org/10.1109/CVPR.2017.571},
  year       = {2017}
}

@inproceedings{worldtree,
  author     = {Peter A. Jansen and
Elizabeth Wainwright and
Steven Marmorstein and
Clayton T. Morrison},
  bibsource  = {dblp computer science bibliography, https://dblp.org},
  booktitle  = {Proceedings of the Eleventh International Conference on Language Resources
and Evaluation, {LREC} 2018, Miyazaki, Japan, May 7-12, 2018},
  editor     = {Nicoletta Calzolari and
Khalid Choukri and
Christopher Cieri and
Thierry Declerck and
Sara Goggi and
K{\^{o}}iti Hasida and
Hitoshi Isahara and
Bente Maegaard and
Joseph Mariani and
H{\'{e}}l{\`{e}}ne Mazo and
Asunci{\'{o}}n Moreno and
Jan Odijk and
Stelios Piperidis and
Takenobu Tokunaga},
  publisher  = {European Language Resources Association {(ELRA)}},
  title      = {WorldTree: {A} Corpus of Explanation Graphs for Elementary Science
Questions supporting Multi-hop Inference},
  url        = {http://www.lrec-conf.org/proceedings/lrec2018/summaries/81.html},
  year       = {2018}
}

@inproceedings{worldtreev2,
  author     = {Zhengnan Xie and
Sebastian Thiem and
Jaycie Martin and
Elizabeth Wainwright and
Steven Marmorstein and
Peter A. Jansen},
  bibsource  = {dblp computer science bibliography, https://dblp.org},
  booktitle  = {Proceedings of The 12th Language Resources and Evaluation Conference,
{LREC} 2020, Marseille, France, May 11-16, 2020},
  editor     = {Nicoletta Calzolari and
Fr{\'{e}}d{\'{e}}ric B{\'{e}}chet and
Philippe Blache and
Khalid Choukri and
Christopher Cieri and
Thierry Declerck and
Sara Goggi and
Hitoshi Isahara and
Bente Maegaard and
Joseph Mariani and
H{\'{e}}l{\`{e}}ne Mazo and
Asunci{\'{o}}n Moreno and
Jan Odijk and
Stelios Piperidis},
  pages      = {5456--5473},
  publisher  = {European Language Resources Association},
  title      = {WorldTree {V2:} {A} Corpus of Science-Domain Structured Explanations
and Inference Patterns supporting Multi-Hop Inference},
  url        = {https://aclanthology.org/2020.lrec-1.671/},
  year       = {2020}
}

@inproceedings{textgraphs,
  author     = {Peter Jansen and
Dmitry Ustalov},
  bibsource  = {dblp computer science bibliography, https://dblp.org},
  booktitle  = {Proceedings of the Thirteenth Workshop on Graph-Based Methods for
Natural Language Processing, TextGraphs@EMNLP 2019, Hong Kong, November
4, 2019},
  doi        = {10.18653/V1/D19-5309},
  editor     = {Dmitry Ustalov and
Swapna Somasundaran and
Peter Jansen and
Goran Glavas and
Martin Riedl and
Mihai Surdeanu and
Michalis Vazirgiannis},
  pages      = {63--77},
  publisher  = {Association for Computational Linguistics},
  title      = {TextGraphs 2019 Shared Task on Multi-Hop Inference for Explanation
Regeneration},
  url        = {https://doi.org/10.18653/v1/D19-5309},
  year       = {2019}
}

@inproceedings{valentino,
  author     = {Marco Valentino and
Mokanarangan Thayaparan and
Andr{\'{e}} Freitas},
  bibsource  = {dblp computer science bibliography, https://dblp.org},
  booktitle  = {Proceedings of the 16th Conference of the European Chapter of the
Association for Computational Linguistics: Main Volume, {EACL} 2021,
Online, April 19 - 23, 2021},
  doi        = {10.18653/V1/2021.EACL-MAIN.15},
  editor     = {Paola Merlo and
J{\"{o}}rg Tiedemann and
Reut Tsarfaty},
  pages      = {200--211},
  publisher  = {Association for Computational Linguistics},
  title      = {Unification-based Reconstruction of Multi-hop Explanations for Science
Questions},
  url        = {https://doi.org/10.18653/v1/2021.eacl-main.15},
  year       = {2021}
}

@article{bge-m3,
  author     = {Jianlv Chen and
Shitao Xiao and
Peitian Zhang and
Kun Luo and
Defu Lian and
Zheng Liu},
  bibsource  = {dblp computer science bibliography, https://dblp.org},
  doi        = {10.48550/ARXIV.2402.03216},
  eprint     = {2402.03216},
  eprinttype = {arXiv},
  journal    = {CoRR},
  title      = {{BGE} M3-Embedding: Multi-Lingual, Multi-Functionality, Multi-Granularity
Text Embeddings Through Self-Knowledge Distillation},
  url        = {https://doi.org/10.48550/arXiv.2402.03216},
  volume     = {abs/2402.03216},
  year       = {2024}
}

@inproceedings{theorem,
  author    = {Max Ku and Cheuk Hei Chong and Jonathan Leung and Krish Shah and Alvin Yu and Wenhu Chen},
  title     = {{TheoremExplainAgent}: Towards Video-based Multimodal Explanations for {LLM} Theorem Understanding},
  booktitle = {Proceedings of the 63rd Annual Meeting of the Association for Computational Linguistics (Volume 1: Long Papers), {ACL} 2025},
  pages     = {6663--6684},
  publisher = {Association for Computational Linguistics},
  year      = {2025},
  doi       = {10.18653/v1/2025.acl-long.332}
}

@article{edustory,
  author  = {Xinyi Wu and Jinesh Teotia and Shuai Zhao and Erik Cambria},
  title   = {{EduStory}: A Unified Framework for Pedagogically-Consistent Multi-Shot {STEM} Instructional Video Generation},
  journal = {CoRR},
  volume  = {abs/2605.09378},
  year    = {2026},
  doi     = {10.48550/arXiv.2605.09378}
}

@article{courseblueprint,
  author  = {Md Zahidul Islam and Md Mahmudul Hasan Manik and Guanghui Wang},
  title   = {{CourseBlueprint}: A Structured Pipeline for Adaptive Pedagogical Video Generation Grounded in Course Corpora},
  journal = {CoRR},
  volume  = {abs/2606.20608},
  year    = {2026},
  doi     = {10.48550/arXiv.2606.20608}
}

@article{eduillustrate,
  author  = {Shaonan Bi and Mingyu Zhang and Zhen Li and Xin Wang and Kun Li and Aimin Zhou},
  title   = {{EduIllustrate}: Towards Scalable Automated Generation of Multimodal Educational Content},
  journal = {CoRR},
  volume  = {abs/2604.05005},
  year    = {2026},
  doi     = {10.48550/arXiv.2604.05005}
}

@article{dpp,
  title     = {Determinantal Point Processes for Machine Learning},
  author    = {Kulesza, Alex and Taskar, Ben},
  journal   = {Foundations and Trends in Machine Learning},
  volume    = {5},
  number    = {2--3},
  pages     = {123--286},
  year      = {2012}
}

\end{document}